\documentclass[aps,prd,notitlepage]{revtex4-1}
\usepackage{subfigure,graphics} 
\usepackage{flafter} 
\begin{document} 

\title{
Covariant one-loop quantum gravity and Higgs inflation} 
\author{Ian G. Moss}
\email{ian.moss@ncl.ac.uk}
\affiliation{School of Mathematics and Statistics, Newcastle University, 
Newcastle Upon Tyne, NE1 7RU, UK}
\date{\today}

%%%%%%%%%%%%%%%%%%%%%%%%%%%%%%%%%%%%%%%%%%%

\begin{abstract}
The quantisation of scalar field theory and Einstein gravity
is investigated  using a fully covariant background field
formalism, including Vilkovisky-DeWitt corrections. 
The one-loop divergences, which 
are relevant for the consistency of the low-energy effective theory,
differ substantially from non-covariant calculations.
The results are applied to the effective action of Higgs inflation, 
with a non-minimal gravity coupling parameter $\xi$, where the
use of a formalism which is independent of field redefinitions
is particularly important. Consistency of the one-loop effective action
requires the cut-off $\Lambda<M_p/\sqrt{\xi}$ in the small field 
and $\Lambda<M_p$ in the large field limit.
\end{abstract}
\pacs{PACS number(s): }
\maketitle

%%%%%%%%%%%%%%%%%%%%%%%%%%%%%%%%%%%%%%%%%%%
\section{introduction}

Quantum gravity is potentially an important ingredient in descriptions of the
very early-universe. Of necessity, most early universe applications of quantum gravity are done 
within the context of an effective field theory, introducing a cut-off scale below which the theory is
described with some degree of accuracy by general relativity coupled to the 
matter sector \cite{Donoghue:1995cz,Burgess:2003jk}.

At energies below the cut-off scale, we are free to calculate the quantities of interest, such as
the effective action, including corrections to the desired order of accuracy. Having relaxed the concept
of renormalisability, we still have to address some important technical questions.
One issue which provides the focus of this paper is how to remove any dependence of the effective action
on the choice of field variables, in other words how to establish covariance on field space.
This is closely linked to obtaining an effective action which is independent of the
choice of gauge-fixing.  This paper will utilise the Vilkovisky-DeWitt method 
\cite{Vilkovisky:1984,DeWitt2} for retaining full covariance, 
focusing on the one loop effective action for the gravity-scalar system.

Since the existence of the cut-off is an important property in the effective theory,
it seems sensible to employ a covariant cut-off regularisation scheme. The Schwinger
method is consistent with gauge invariance and general covariance and is
well-suited to evaluation of the one loop effective action
\cite{barvinsky1985generalized,avramidi2000heat,vassilevich2003heat}. The divergences become
finite, cut-off dependent terms which are important for analysing the self consistency of the
theory. These terms are  also very closely related to the ones needed in the search for 
renormalisation group fixed points, which may provide special classes of 
ultraviolet-complete effective theories if the asymptotic-safety philosophy is adopted
\cite{Weinberg:1977}.

The first non-trivial type of divergence is the quadratic divergence proportional to $\Lambda^2$. 
The quadratic divergences appear in many places, for example as corrections to the effective 
Planck mass. This gives the first example of a result which potentially
depends on the gauge fixing terms, and the covariant approach gives a new
result for the running of the Planck mass,
\begin{equation}
M_p^2(\Lambda)=M_p^2(0)+{\Lambda^2\over 24\pi^2}N,
\end{equation}
where $N$ is the number of scalar fields. 
The mass $M_p(0)$ can be regarded as the cosmological Planck mass
whilst $M_p(\Lambda)$ is the effective mass for small-scale quantum fluctuations.
The covariant approach gives a running part to the mass which has has the opposite sign to 
the one found in previous non-covariant calculations \cite{Larsen:1995ax,Calmet:2008tn}.

The second part of this paper considers the importance of quantum corrections
to the cosmological model of Higgs inflation
\cite{Futamase:1987ua,Salopek:1988qh,Fakir:1990eg,Kaiser:1994vs,Komatsu:1999mt,Bezrukov:2007ep}.
In Higgs inflation, the Standard Model of particle
physics is coupled to gravity with the Einstein term and a non-minimal Higgs
coupling $\xi R|{\cal H}|^2$. This non-minimal coupling term can be removed by
a conformal re-scaling of the metric, the original metric defining the Jordan frame and the
new metric defining the Einstein frame. The Jordan frame has the simpler action, but
Einstein frame is convenient for descriptions of inflation. 

Quantum corrections to Higgs inflation were first considered with reference to the
running of the Higgs self-coupling 
\cite{DeSimone:2008ei,Bezrukov:2008ej,Barvinsky:2008ia,Barvinsky:2009fy,Barvinsky:2010yb}. 
This affects the shape of the Higgs potential in the
inflationary regime, and through this predictions of the large-scale structure
of the universe. Since the running of the Higgs self-coupling is sensitive
to low energy physics, there seemed to be an interesting new link between particle
physics and cosmology. However, there where early indications,
based on power counting in Feyman diagrams, that the 
quantum corrections to the model became problematic if the cut-off is at an energy scale 
around $M_p/\xi$, which would be below the inflationary scale in the Einstein frame  
\cite{DeSimone:2008ei,Burgess:2009ea,Barbon:2009ya}. The consistency seemed
at first to be restored if the quantum theory was done in the Einstein frame
\cite{Lerner:2009na}, but then problems arose with the Goldstone boson sector
even in the Einstein frame \cite{Burgess:2010zq,Hertzberg:2010dc}.
Subsequently, background field techniques have indicated that the limit $M_p/\xi$ applies in the 
small field regime but becomes $M_p$ in the large field regime relevant to
inflation \cite{Bezrukov:2010jz,George:2013iia,Steinwachs:2013tr,Prokopec:2014iya}.

In order to be able to obtain identical results in both the Jordan and Einstein frames it seems
desirable to employ a quantum technology which is covariant under field definitions,
hence the relevance of the Vilkovisky-DeWitt methodology
(as has been suggested in \cite{Steinwachs:2013tr}). Furthermore, since ghost loops can affect
the degree of divergence in simple power counting arguments, a full one-loop
analysis will be attempted in Section \ref{higgs}.

The literature on one-loop gravity is substantial. Some of the early work on gravity-scalar
systems was done by DeWitt \cite{DeWittdynamical,DeWitt67} and t'Hooft and Veltman \cite{thooft:1974}. 
Reviews and further references can be found in 
\cite{barvinsky1985generalized,avramidi2000heat,vassilevich2003heat,ParkerTomsbook}.
Results which use Vilkoviski-DeWitt corrections include gauge couplings
and the Higgs mass corrections in \cite{He:2010mt} and scalar
kinetic and mass terms in \cite{Mackay:2009cf}.
The gauge-fixing functionals used here are based on one first used by DeWitt, and supplemented
with scalar terms by  t'Hooft and Veltman (not to be confused with their electromagnetic gauge). 
In this paper, Feynman gauge refers to gauge-fixing with gauge parameter
$\alpha=1$ and Landau gauge to gauge fixing with $\alpha\to0$. The metric conventions used
in this paper follow Misner, Thorne and Wheeler \cite{MTW} and the units are
ones in which $c=\hbar=1$.

\section{One-loop calculations}

In the Vilkovisky-DeWitt formalism, the one-loop contribution to the effective action
consists of a contribution from the original fields $\varphi^i$ and a contribution
from ghosts $c_i$. The field contribution is given in condensed notation by
\begin{equation}
\frac{i}2\lim_{\alpha\to 0}
\ln\det\left(\nabla^i\nabla_jS[\varphi]
+{1\over 2\alpha}K^i{}_\epsilon[\varphi]K_j{}^\epsilon[\varphi]\right),
\label{vdw}
\end{equation}
where the indices $i,j$ represent both coordinate and internal components.
The innovation of Vilkovisky and DeWitt was to put the second functional derivatives into covariant form,
\begin{equation}
\nabla_i\nabla_j S=S[\varphi]_{,ij}-\Gamma^k{}_{ij}S[\varphi]_{,k},
\end{equation}
where the connection coefficients $\Gamma^i{}_{jk}$ will be determined by the Levy-Civita connection for 
the metric on the space of fields. The connection ensures that the result is covariant under 
field redefinitions. It can be disregarded when the background field is on-shell, i.e. $S_{,i}=0$, 
but any off-shell application of the effective action has to include it.

The second term in (\ref{vdw}) is a gauge-fixing term for a gauge-fixing functional
$\chi_\epsilon=K^i{}_\epsilon\delta\varphi_i$, which uses the generator of 
gauge transformations $K^i{}_\epsilon$. Other gauge-fixing terms can be used,
without changing the form of the effective action, provided the field-space connection
is suitably modified. However, the one-loop effective action
obtained from the Landau gauge in (\ref{vdw}) 
is identical to the  one-loop effective action which is fully independent of 
the gauge-fixing term.

\subsection{Quadratic action}

The first task is to evaluate the functional derivatives appearing the
one-loop determinant. The basic fields under consideration are the spacetime metric 
$g_{\mu\nu}$, with Ricci scalar $R$, and a set of scalar fields $\phi^i$. 
The scalar fields will take values in an internal manifold,
with an internal metric ${\cal G}_{ij}$ and potential $V$.
The number of spacetime dimensions is $m$ and the internal space dimension
is  N. The Lagrangian densities for gravity ${\cal L}_g$, 
scalar ${\cal L}_s$ and gauge fixing ${\cal L}_\chi$ are,
\begin{eqnarray}
{\cal L}_g&=&{1\over 2\kappa^2}R\,|g|^{1/2},\\
{\cal L}_s&=&-\frac12 {\cal G}_{ij}(\phi)\,g^{\mu\nu}\partial_\mu\phi^i\partial_\nu\phi^j\,|g|^{1/2}
-V(\phi)\,|g|^{1/2},\label{nlsm}\\
{\cal L}_\chi&=&-\alpha^{-1}\chi^\mu\chi_\mu,
\end{eqnarray}
where $\partial_\mu$ denotes an ordinary spatial derivative.
These are expanded about a background field configuration, so that 
${\cal L}={\cal L}_0+{\cal L}_1+\dots+J^\mu{}_{,\mu}$, where
${\cal L}_n$ is $n$'th order in perturbations and $J^\mu$ is chosen so 
that all the terms are first order in derivatives.

We can expand the metric in a simple linear fashion about a background $g_{\mu\nu}$,
making the replacement  $g_{\mu\nu}\to g_{\mu\nu}+2\kappa\gamma_{\mu\nu}$.
Expansion of the Einstein action to quadratic order has been descibed in many places
\cite{barvinsky1985generalized,avramidi2000heat,vassilevich2003heat,ParkerTomsbook},
and the results are combined in Appendix \ref{ap}.
The scalar field is best expanded using the covariant background field method
\cite{DeWitt67,AlvarezGaume198185} where $\eta^i$
is the tangent vector to the (internal space) geodesic joining the background
field $\varphi^i$ to $\phi^i$. The covariant background field approach removes the need to include
the connection terms for the scalars in Eq. (\ref{vdw}), and instead we include them in the
derivatives of $\eta^i$,
\begin{equation}
D_\mu\eta^i=\nabla_\mu\eta^i+\partial_\mu\varphi^k\Gamma^i{}_{jk}\eta^j.
\end{equation}
The curvature tensor on the internal space will be denoted by ${\cal R}_{ijkl}$.
The quadratic order Lagrangian densities are given explicitly in Appendix \ref{ap}.

The gauge transformations are $\delta_c\gamma_{\mu\nu}=c_{\mu;\nu}+c_{\nu;\mu}$ and
$\delta_c\eta^i=2\kappa c^\mu\partial_\mu\varphi^i$.
These are generated by the gauge-fixing functional of the form,
\begin{equation}
\chi^\mu=g^{(\mu\nu)(\rho\sigma)}\gamma_{\rho\sigma;\nu}-\kappa\,\partial^\mu\phi^i\,\eta_i,
\end{equation}
where $g^{(\mu\nu)(\rho\sigma)}$ is the DeWitt metric,
\begin{equation}
g^{(\mu\nu)(\rho\sigma)}=\frac12\left(g^{\mu\rho}g^{\nu\sigma}+g^{\mu\sigma}g^{\nu\rho}
-g^{\mu\nu}g^{\rho\sigma}\right).
\end{equation}
The effective action also has the ghost contribution, obtained from the gauge variation of the gauge-fixing
functional $\chi^\mu$ under the diffeomorphism symmetry.

The final ingredient in  (\ref{vdw}) is the set of field-space connection terms and background
field equations. These have been evaluated 
on flat spacetime backgrounds by Mackay and Toms \cite{Mackay:2009cf}, but they generalise to curved 
backgrounds very simply. The connection coefficients can be evaluated using standard formulae
for the Levy-Civita connection with the field-space metrics, 
which in the case of gravitational and scalar perturbations are 
$g^{(\mu\nu)(\rho\sigma)}\,|g|^{1/2}\delta(x-x')$ 
and ${\cal G}_{ij}|g|^{1/2}\delta(x-x')$.
For example, there is a contribution to the scalar sector from the field-space connection coefficient
(dropping the $x$ dependence),
\begin{equation}
\Gamma^{(\mu\nu)}{}_{ij}={\kappa\over m-2}g_{\mu\nu}{\cal G}_{ij}.
\end{equation}
The background field equation in this case is the Einstein field equation 
$G_{\mu\nu}-\kappa^2 T_{\mu\nu}=0$.
The full set of Vilkovisky-DeWitt corrections, including this one,  is given in Eq. (\ref{Lv})

When the various contributions are combined it makes sense to write the result in a form which
is suited to the operator methods which will be used to evaluate the effective action. We can
combine the metric and scalar variations into a single bosonic field $\Phi$,
\begin{equation}
\Phi_a=\pmatrix{\gamma_{\mu\nu}\cr\eta_i}.
\end{equation}
It is also advantageous to rewrite the terms involving a single spacetime derivative in terms of
an effective gauge potential ${\cal A}_\mu$, and then
\begin{eqnarray}
{\cal L}_2\,|g|^{-1/2}&=&-\frac12{\cal G}^{ab}{\cal D}_\mu\Phi_a{\cal D}^\mu\Phi_b
-\frac12\zeta\,{\cal P}^{\mu\nu ab}{\cal D}_\mu\Phi_a{\cal D}_\nu\Phi_b
-\frac12{\cal M}^{2\,ab}\Phi_a\Phi_b\nonumber\\
&&-{\cal G}^{ab}\Phi_a{\cal A}_\mu{}_a{}^b{\cal D}^\mu\Phi_b
+{\cal G}^{ab}{\cal D}^\mu\Phi_a{\cal A}_\mu{}_a{}^b\Phi_b,\label{Leasy}
\end{eqnarray}
where ${\cal D}_\mu$ is the matrix of spacetime covariant derivatives. Indices are raised using the 
metric ${\cal G}^{ab}$, where
\begin{equation}
{\cal G}^{ab}=\pmatrix{g^{(\mu\nu)(\rho\sigma)}&0\cr0&{\cal G}^{ij}\cr}
\end{equation}
The tensor ${\cal P}^{\mu\nu ab}$ comes from the gravitational sector (see \ref{ptens}),
\begin{equation}
{\cal P}^{\alpha\beta ab}=\pmatrix{{\cal P}^{\alpha\beta(\mu\nu)(\rho\sigma)}&0\cr0&0\cr}
\end{equation}
The effective gauge potential mixes the gravity and scalar sectors,
\begin{equation}
{\cal A}_{\alpha\,a}{}^b=\pmatrix{0&{\cal A}_{\alpha(\mu\nu)}{}^j\cr 
{\cal A}_{\alpha i}{}^{(\rho\sigma)}&0\cr},
\end{equation}
where 
${\cal A}_{\alpha\,(\mu\nu)}{}^i=-{\cal A}_\alpha{}^i{}_{(\mu\nu)}
=\zeta\kappa g_{\alpha(\mu}\partial_{\nu)}\phi^j$
can be read off from Eqs. (\ref{Ls}) and (\ref{Lchi}).

\subsection{Regularisation and divergences}

The proper-time cut-off regularisation scheme devised by Schwinger proves a fully 
gauge-invariant means of regularising the one-loop effective action, and gives an explicit 
representation for the divergent terms at each order in the cut-off
\cite{avramidi2000heat,vassilevich2003heat}. The method is
defined in terms of the heat kernel $K(x,x',\tau)$ of an elliptic operator 
$\Delta$,
\begin{equation}
K(x,x',\tau)=\sum_n u_n(x)u^\dagger_n(x')e^{-\lambda_n\tau},
\end{equation}
where $u_n(x)$ are the normalised eigenfunctions of $\Delta$ with eigenvalues $\lambda_n$.
In order to use the proper-time method it is necessary to find an analytic continuation
of the spacetime from the metric signature $-+++$ to an elliptic signature $++++$.

The definition of $\ln\det\Delta$ with proper-time cut-off $\tau$ is provided by
\begin{equation}
\ln\det\Delta=-\int d^mx\,|g|^{1/2}\int_\tau^\infty\,{d\tau'\over \tau'} {\rm Tr}\,K(x,x,\tau'),
\label{defreg}
\end{equation}
where ${\rm Tr}$ is over the internal indices. For a second order operator, the behaviour of 
the heat kernel for small $\tau$ is determined by an asymptotic expansion,
\begin{equation}
K(x,x,\tau)\sim (4\pi \tau)^{-m/2}\sum_{r=0}^\infty E_r(\Delta,x)\tau^r.
\end{equation}
The trace of each coefficient $E_r(x)$ for a covariant operator is given by a local expression 
invariant under the gauge symmetries.

The small $\tau$ expansion can be used to isolate the divergent parts of $\ln\det\Delta$.
Dimensionally,  $\tau=1/\Lambda^2$, where $\Lambda$ is an energy scale. 
The divergent part of $\ln\det\Delta$ expressed in terms of $\Lambda$ is then
\begin{equation}
\left.\ln\det\Delta\right|_{\rm div}\sim- 
{1\over (4\pi)^{m/2}}\int d^4 x\,|g|^{1/2}\left(\sum_{r=0}^{m-1}
{2\over m-2r} {\rm Tr}\,E_r(\Delta,x)\Lambda^{m-2r}
+{\rm Tr}\,E_{m/2}\ln\Lambda\right).\label{dp}
\end{equation}
The divergent part of the one-loop effective action $\Gamma_{\rm div}$ is has contributions from
the  fields with operator $\Delta_f$ and the ghosts with operator $\Delta_g$,
\begin{equation}
\Gamma_{\rm div}=\lim_{\alpha\to 0}\left\{
-\frac12\left.\ln\det\Delta_f\right|_{\rm div}
+\left.\ln\det\Delta_g\right|_{\rm div}
\right\}.\label{gdiv}
\end{equation}
Analytic continuation back to the metric signature $-+++$ has introduced a factor `$i$' from the
spacetime volume integration. In flat spacetime, the cut-off $\Lambda$ becomes the usual
energy-momentum cut-off, but in curved space-time there may be a need to re-scale $\Lambda$
to a physical cut-off relevant to an observer's frame of reference.

Explicit expressions for the traces ${\rm Tr}\,E_n(x)$ are known for certain operators
\cite{avramidi2000heat,vassilevich2003heat}.
The relevant operator for the Einstein-scalar system can be read off from the
Lagrangian density (\ref{Leasy}).
There are limited results on this type of non-minimal operator, but recent progress has been made
on ${\rm Tr}\,E_1$ and ${\rm Tr}\,E_2$ \cite{Moss:2013cba}. The general result is known for ${\rm Tr}\,E_1$,
and is given in Appendix \ref{appb}. Partial results for ${\rm Tr}\,E_2$ are given in
Appendix \ref{appc}.

In four dimensions, the ${\rm Tr}\,E_1$ terms are the coefficients of quadratic divergences. 
From this point on the value of $m$ will be fixed at $m=4$ and the reduced Planck mass
$M_p=1/\kappa$. The results of Appendix \ref{appb} translate into the quadratic divergences
$\Gamma_{\rm quad}$,
\begin{eqnarray}
\Gamma_{\rm quad}&=&{1\over 32\pi^2}{\Lambda^2\over M_p^2}
\int d^4x|g|^{1/2}\left\{-{N+22\over 3}M_p^2\,R+
\frac12N\partial_\mu\varphi^i\partial^\mu\varphi_i\right.\nonumber\\
&&\left.+2(N+6)V+M_p^2\,{\cal R}_{ij}\partial_\mu\varphi^i\partial^\mu\varphi^j
-M_p^2\,V_{;i}{}^i\right\}\label{qd}
\end{eqnarray}
Recall that we started from an effective theory in which the cut-off takes a finite
value, representing some high-energy physics which has been removed from the
problem, therefore $\Gamma_{\rm quad}$ is a finite contribution to the effective action. 

The coefficient of the curvature term has a part depending on $N$, which comes from the
scalar sector. However, the magnitude and even the sign of this contribution is
different from previous scalar calculations because of the Vilkovisky-DeWitt
corrections to the effective action 
\cite{Larsen:1995ax,Calmet:2008tn,Atkins:2010re}. The new result differs by
factors of $R+\kappa^2T$, which vanish on-shell but which affect
the renormalisation in a critical way. If we interpret the original action as an effective
theory, then the bare mass $M_p$ is the value obtained from integrating out unknown physical
effects from energy scales above $\Lambda$, therefore the bare mass depends on $\Lambda$. The 
Planck mass relevant for cosmology is taken from the coefficient of the Ricci scalar
in the effective action, which we call $M_p^2(0)$, giving the scaling relation, 
\begin{equation}
M_p^2(\Lambda)=M_p^2(0)+{1\over 24\pi^2}\Lambda^2(N+22).
\end{equation}
The scalar kinetic and potential terms can be partly absorbed by field and mass
renormalisations, but for ${\cal R}_{ijkl}\ne 0$ the effective action also contains terms
which have to be included in the effective field theory. The self-consistency of the
one-loop action depends on these new terms being small, hence there is a
requirement that $\Lambda\ll M_p(\Lambda)$.

Current technology is not able to deliver the general expression for ${\rm Tr}\,E_2$ for the 
full Einstein-scalar operator, so the discussion of the logarithmic divergences will be restricted
to constant backgrounds. $\partial_\mu\varphi^i=0$. The logarithmic
divergences from Eq. (\ref{dp})  are
\begin{eqnarray}
\Gamma_{\rm log}&=&{1\over 32\pi^2}\ln{\Lambda}
\int d^4x|g|^{1/2}\left\{
\frac12V_{;ij}V^{;ij}-2M_p^{-2}VV_{;i}{}^i\right.\nonumber\\
&& -\frac32M_p^{-2}V_{;i}V^{;i}+(2N+12)M_p^{-4}V^2\nonumber\\
&&+\frac13RV_{;i}{}^i-{2N+24\over 3}M_p^{-2}RV\nonumber\\
&&\left.+\frac{N+212}{180}R_{\mu\nu\rho\sigma}^2-\frac{N+122}{180}R_{\mu\nu}^2
+{2N+25\over36}R^2
\right\}.\label{ld}
\end{eqnarray}
The first term comes from the scalar sector and includes the conventional
renormalisation of the scalar quartic coupling constant. The other terms
have a contribution from the metric fluctuations, and their coefficients
depend on the inclusion of Vilkovisky-DeWitt corrections. The effective theory
must have terms at least up to order $\varphi^8$ in the potential to absorb the
dependence on $\Lambda$. The non-minimal coupling terms such as
$RV_{;i}{}^i$ can removed by conformal re-scalings, as we shall see in the
next section.

Contributions of the running of the coupling constants in the Standard Model of
particle physics suggest that the effective Higgs self-coupling may become negative at large 
values of the Higgs field \cite{Degrassi:2012ry}. This de-stabilising effect may be 
overcome by higher order terms in the Higgs
potential which arise from Higgs-gravity interactions.
These terms are undetermined in the effective theory, but may have some definite
coefficients if the large-scale behaviour is determined by a (Wilsonian style)
renormalisation group fixed point.

\section{Higgs Inflation}\label{higgs}

The Lagrangian density for Higgs inflation in the Jordan frame contains the standard
model Higgs terms and a non-minimal
coupling between the metric $\hat g_{\mu\nu}$ and the Higgs doublet field ${\cal H}$.
The gravity-scalar contributions are
\begin{equation}
{\cal L}|\hat g|^{-1/2}={1\over 2\kappa^2}\hat R-\xi \hat R\,{\cal H}^\dagger{\cal H}
-\hat g^{\mu\nu}(D_\mu{\cal H})^\dagger(D_\mu{\cal H})
-\hat V({\cal H}).
\end{equation}
The Lagrangian density in the Einstein frame is obtained by the conformal transformation
\begin{equation}
\hat g_{\mu\nu}=f\,g_{\mu\nu},
\end{equation}
where
\begin{equation}
f=(1+2\kappa^2\xi{\cal H}^\dagger{\cal H})^{-1}
\end{equation}
Total derivatives terms have to be removed by integrating by parts, and then
\begin{equation}
{\cal L}|g|^{-1/2}={1\over 2\kappa^2}R
-f(D_\mu{\cal H})^\dagger(D^\mu{\cal H})
-\frac32\xi^2\kappa^2 f^2\partial_\mu({\cal H}^\dagger{\cal H})
\partial^\mu({\cal H}^\dagger{\cal H})
-f^2\hat V({\cal H}),
\end{equation}
where indices are raised using the Einstein metric. The scalar part of the action is a non-linear 
sigma model similar to (\ref{nlsm}) when written in terms of the real components 
${\cal H}^T=(\phi^1+i\phi^2,\phi^3+i\phi^4)/\sqrt{2}$, with $V=f^2\hat V$ and
\begin{equation}
{\cal G}_{ij}=f\delta_{ij}+6f^2\xi^2\kappa^2\delta_{ik}\delta_{jl}\phi^k\phi^l.
\end{equation}
When evaluated at the background field $\varphi^i$, the metric can be expressed in 
terms of a canonically normalised massive boson and transverse Goldstone mode directions,
\begin{equation}
{\cal G}_{ij}=f(\varphi)\delta_{ij}^{\perp}+\chi_{,i}\chi_{,j}
\end{equation}
where $\varphi^2=\delta_{ij}\varphi^i\varphi^i$, $f(\varphi)=(1+\xi\kappa^2\varphi^2)^{-1}$
 and 
\begin{equation}
\chi=\int d\varphi f(\varphi)\left[1+(6\xi+1)\xi\kappa^2\varphi^2\right]^{1/2}
\end{equation}
The Ricci curvature and potential derivative terms for the metric are
\begin{eqnarray}
{\cal R}_{\chi\chi}&=&-\frac32f^{-1}f_{,\chi\chi}-\frac34f^{-2}f_{,\chi}^2,\\
{\cal R}_{ij}^\perp&=&-\left(\frac12f_{,\chi\chi}+\frac34f^{-1}f_{,\chi}^2\right)\delta_{ij}^{\perp},\\
V_{;i}{}^i&=&(f^2\hat V)_{,\chi\chi}+\frac32f^{-1}f_{,\chi}(f^2\hat V)_{,\chi}.\label{gradv}\\
V_{;ij}V^{;ij}&=&(f^2\hat V)_{,\chi\chi}^2
+\frac34f^{-2}f_{,\chi}^2(f^2\hat V)_{,\chi}^2.\label{gradvs}
\end{eqnarray}
When $\xi$ is large, the derivative of the potential is small for large values of the 
canonical field $\chi$ rendering this regime suitable for inflation.

The Jordan and Einstein frames are related by a field redefinition, and the
effective action calculated using the covariant approach on field space will be the same, 
in both frames apart form the definition of the cut-off $\Lambda$. In the proper-time
regularisation scheme, the relation between the proper-times and the cosmological times
changes due to the conformal transformation between the frames, hence  
the cut-off in the Jordan frame $\hat\Lambda$ and the Einstein frame $\Lambda$ are related by
\begin{equation}
\hat\Lambda=f^{-1/2}\Lambda.
\end{equation} 
The quadratic divergences for the Einstein-scalar sectors of the theory in the Einstein frame are given
by the earlier result (\ref{qd}). Consider the kinetic terms in the Lagrangian density of the Higgs boson $\chi$,
\begin{equation}
-\frac12{\cal G}_{\chi\chi}\partial_\mu\chi\partial^\mu\chi.
\end{equation}
The background value ${\cal G}_{\chi\chi}=1$, and the quadratic divergences contribute a correction,
\begin{equation}
{\cal G}_{\chi\chi}{}_{\rm quad}=--{1\over4\pi^2}{\Lambda^2\over M_p^2}
-{1\over 16\pi^2}\Lambda^2{\cal R}_{\chi\chi}
\end{equation}
It is worth recalling that the ${\cal R}_{\chi\chi}$ term originates from the background
 mass term ${\cal R}_{i\chi}{}^j{}_\chi$, where $i$ and $j$ are in the transverse, or Goldstone
boson directions, therefore this term is a Goldstone boson loop contribution.
In the large and small $\varphi$ limits the curvature becomes
\begin{equation}
R_{\chi\chi}=\cases{\displaystyle
-{3\xi\over M_p^2}&$\varphi\ll M_p/\sqrt{\xi}$\cr
&\cr
\displaystyle
-{5\over 6 M_p^2}&$\varphi\gg M_p/\sqrt{\xi}$\cr
}
\end{equation}
The self-consistency of the effective theory requires that 
$|{\cal G}_{\chi\chi}{}_{\rm quad}|<1$, setting an upper limit on the cut-off scale
of $\Lambda<M_p/\sqrt{\xi}$ for small $\varphi$ and $\Lambda<M_p$
for large $\varphi$. The small $\varphi$ limit is slightly weaker than the result
obtained by power counting \cite{Bezrukov:2010jz}, but the large $\varphi$ 
limit is identical.

The regime of interest for Higgs inflation is the large field limit, where the potential 
has the standard model form in the Jordan frame. In this regime we can ignore the
Higgs mass terms and use the potential $\hat V=\lambda|{\cal H}|^4$. 
In the Einstein frame, the potential of the background field $\varphi$ becomes
\begin{equation}
V(\varphi)=\frac14\lambda f(\varphi)^2\varphi^4
\sim\frac14{\lambda\over\xi^2}M_p^4-\frac12{\lambda\over\xi^3}{M_p^6\over\varphi^2}
+\dots
\end{equation}
The quadratic divergence contributes a term $V_{\rm quad}$ (in the $\xi\gg1$ limit),
which can be read off from the effective action Eq. (\ref{qd}). The $V$ term is absorbed 
by a renormalisation of the quartic coupling, leaving the term (\ref{gradv}),
\begin{equation}
V_{\rm quad}={1\over 32\pi^2}\Lambda^2V_{;i}{}^i\sim
-{3\over 16\pi^2}{\lambda\over \xi^3}{\Lambda^2M_p^4\over\varphi^2}+\dots
\end{equation}
The leading term in the potential is therefore unaffected by these particular quantum corrections.
However, the slope of the potential is important for the observational predictions
of inflation, and keeping the contribution from the quadratic divergences small requires a cut-off scale 
$\Lambda<M_p$. The coefficient of $\varphi^{-2}$ has to be regarded as a new parameter
in the effective theory with an undetermined value, independent of the quartic Higgs coupling. 
The slope of the potential is therefore decoupled from the low energy physics of the Higgs boson. 
This agrees with the conclusion reached in \cite{Bezrukov:2010jz}, which was based on examination 
of the $\chi^6$ vertex. We can be confident that the new one-loop result is frame independent and 
fully gauge invariant because the calculation has been done using the formalism which is independent of the 
choice of field variables.

The logarithmic divergences in Eq. (\ref{ld}) do not change the conclusions of the one-loop
calculations in any substantial way. For example, consider the term
\begin{equation}
{1\over 32\pi^2} V_{;ij}V^{;ij}\,\ln\Lambda \sim {288\over\pi^2}{\lambda^2\over\xi^6}
{M_p^8\over\varphi^4}\ln\Lambda
\end{equation}
This is smaller than the original term in the potential by a factor of order $\xi^{-2}$.

\section{Conclusions}

The one-loop effective action for an effective theory of gravity coupled to scalar
field has been analysed using an approach which is covariant under field
redefinitions and independent of gauge-fixing terms, leading to new results
on the running of the coupling constants in the effective theory. When applied to
Higgs inflation, the consistency bounds on the cut-off scale are broadly in line
with recent results obtained using power counting in Feynman diagrams. 

The new results open up the possibility of analysing the full Higgs potential
with one-loop quantum gravity contributions. There are limitations set by
the necessity for introducing undetermined new terms in the Higgs potential
and the undetermined cut-off scale. However, (Wilsonian-style) renormalisation 
group flows can give further information on the existence of possible fixed point 
theories \cite{Reuter:2012xf,Codello2009414}, which may be important if
no new physics intervenes at very large energies.
These renormlisation group
should be addressed using a covariant formalism, and this can be done using heat
kernel coefficients similar to the ones used here.

The new results should also be taken into account when analysing
the effects  of renormalisation group flows on inflation. The scale dependence of 
physical parameters can affect the cosmological predictions. For example, 
the effective Planck mass for large scale cosmology can be different from the 
effective Planck mass for quantum fluctuations. Furthermore,
fixed-point theories are interesting candidates for cosmological models
\cite{Weinberg:2009wa,Hindmarsh:2011hx,Xianyu:2014eba}. 

Finally, the one-loop results obtained here can be improved in a number of ways.
The logarithmic terms can be extended to analyse the $(\partial\varphi)^4$ 
and $R^2(\partial\varphi)^2$ terms, although the calculation would be very demanding. It should also be 
possible to obtain results for the full covariant one-loop effective action, and not just the divergent parts,
on space-time backgrounds of interest to early-universe cosmology.

\acknowledgments

The author is partially supported by  the UK
Science and Technology Facilities Council Consolidated  Grant
ST/J000426/1. He would like to thank the ``Centro de Ciencias Benasque 
Pedro Pascual'' (Spain), for its hospitality during the writing of this paper.

\appendix
\section{Quadratic actions}\label{ap}

This appendix contains the quadratic actions obtained after the expansions described in the
main body of the paper. Most of these results are standard, however most accounts omit the
Vilkovisky-DeWitt corrections. The various contributions to the quadratic action are 
gravity ${\cal L}_g$, scalar ${\cal L}_s$,
gauge-fixing ${\cal L}_\chi$, Vilkovisky-DeWitt corrections ${\cal L}_v$ and
ghosts ${\cal L}_c$ given by
\begin{eqnarray}
{\cal L}_{2g}\,|g|^{-1/2}&=&
-\frac12 g^{(\mu\nu)(\rho\sigma)}\gamma_{\mu\nu;\alpha}\gamma_{\rho\sigma}{}^{;\alpha}
+\frac12 P^{\alpha\beta(\mu\nu)(\rho\sigma)}\gamma_{\mu\nu;\alpha}\gamma_{\rho\sigma;\beta}\nonumber\\
&&+\frac12\left(2R^{\mu\rho\nu\sigma}-g^{\mu\nu}R^{\rho\sigma}
-g^{\rho\sigma}R^{\mu\nu}+2g^{\mu\rho}R^{\nu\sigma}
-g^{(\mu\nu)(\rho\sigma)}\right)\gamma_{\mu\nu}\gamma_{\rho\sigma},\label{Lg}\\
{\cal L}_{2s}\,|g|^{-1/2}&=&-\frac12 D_\mu\eta^iD^\mu\eta_i+\frac12\partial_\mu\varphi^i\partial^\mu\varphi^j
{\cal R}_{ikjl}\eta^k\eta^l-\frac12V_{;ij}\eta^i\eta^j\nonumber\\
&&-\kappa^2g^{(\mu\nu)(\rho\sigma)}g_{\alpha\nu}\gamma_{\rho\sigma}\gamma_{\mu\beta}
\left(T^{\alpha\beta}-\partial^\alpha\varphi^i\partial^\beta\varphi_i\right)
-\kappa g^{\mu\nu}\gamma_{\mu\nu}V_{,i}\eta^i\nonumber\\
&&+\kappa\,g^{(\mu\nu)(\rho\sigma)}\left(
\gamma_{\rho\sigma}\partial_\mu\phi^iD_\nu\eta_i-\gamma_{\rho\sigma;\nu}\partial_\mu\phi^i\eta_i
+\gamma_{\rho\sigma}\phi^i{}_{;\mu\nu}\eta_i\right),\label{Ls}\\
{\cal L}_\chi\,|g|^{-1/2}&=&-\frac12\alpha^{-1}
P^{\alpha\beta(\mu\nu)(\rho\sigma)}\gamma_{\mu\nu;\alpha}\gamma_{\rho\sigma;\beta}
-\alpha^{-1}\kappa^2\,\partial_\mu\phi^i\partial^\mu\phi^j\eta_i\eta_j\nonumber\\
&&-\alpha^{-1}\kappa\,g^{(\mu\nu)(\rho\sigma)}\left(
\gamma_{\rho\sigma}\partial_\mu\phi^iD_\nu\eta_i-\gamma_{\rho\sigma;\nu}\partial_\mu\phi^i\eta_i
+\gamma_{\rho\sigma}\phi^i{}_{;\mu\nu}\eta_i\right),
\label{Lchi}\\
{\cal L}_v\,|g|^{-1/2}&=&-\kappa^2g^{(\mu\nu)(\rho\sigma)}g_{\alpha\nu}\gamma_{\rho\sigma}\gamma_{\mu\beta}
\left(G^{\alpha\beta}-\kappa^2T^{\alpha\beta}\right)+
\frac12(m-2)^{-1}g^{(\mu\nu)(\rho\sigma)}\gamma_{\mu\nu}\gamma_{\rho\sigma}(G-\kappa^2T)\nonumber\\
&&-\frac12\kappa g^{\mu\nu}\gamma_{\mu\nu}\left(\phi^i{}_{;\mu}{}^\mu-V^{;i}\right)\eta_i
+\frac12(m-2)^{-1}(G-\kappa^2T)\eta^i\eta_i,\label{Lv}\\
{\cal L}_c\,|g|^{-1/2}&=&-\frac12c_{\mu;\alpha}c^{\mu;\alpha}+\frac12 R^{\mu\nu}c_\mu c_\nu
+2\kappa^2\,\partial^\mu\varphi^i\partial^\nu\varphi_i\,c_\mu c_\nu.\label{Lc}
\end{eqnarray} 
where the tensor $P$ is
\begin{equation}
P^{\alpha\beta(\mu\nu)(\rho\sigma)}=
2g_{\gamma\delta}g^{(\alpha\gamma)(\rho\sigma)}g^{(\beta\delta)(\rho\sigma)}.\label{ptens}
\end{equation}
The gravity and gauge-fixing contributions involving $P$ combine with coefficient
\begin{equation}
\zeta=\alpha^{-1}-1
\end{equation}
This has lead to most work on quantum gravity making use of the simple case $\alpha=1$, 
sometimes called Feynman or Feynman-DeWitt gauge. However, the effective action generally 
depends on the gauge parameter $\alpha$,  unless the covariant approach is adopted in which case
the correct result is equivalent to the Landau gauge $\alpha\to0$ limit.

Some terms in the scalar part of the Lagrangian density have been simplified by introducing the 
scalar stress-energy tensor,
\begin{equation}
T_{\mu\nu}=\partial_\mu\varphi^i\partial_\nu\varphi_i-
g_{\mu\nu}\left(\frac12\partial_\alpha\varphi^i\partial^\alpha\varphi_i+V\right)
\end{equation}
This highlights a cancellation between the scalar part and the Vilkovisky-DeWitt corrections,
which actually goes further and includes the cancellation of some Einstein tensor terms.

\section{The $E_1$ heat kernel coefficient}\label{appb}

The field operator for the Einstein-scalar system is
\begin{equation}
\Delta_f=-{\cal D}_\alpha{\cal D}^\alpha-
\zeta {\cal P}^{\alpha\beta}{\cal D}_\alpha{\cal D}_\beta
-{\cal A}_\alpha{\cal D}^\alpha-{\cal D}^\alpha{\cal A}_\alpha+{\cal M}{}^2,\label{op}
\end{equation}
The non-minimal part of the operator is governed by the tensor $P^{\alpha\beta}$
in (\ref{ptens}).
This tensor has two useful properties, firstly the symbol
$\sigma(P)={\cal P}^{\alpha\beta}k_\alpha k_\beta/k^2$ satisfies the identity $\sigma(P)^2=\sigma(P)$
and secondly ${\cal D}_\alpha{\cal P}^{\beta\gamma}=0$. Note that combining the gauge potential
with the covariant derivative ${\cal D}_\mu$ would violate the latter property because 
$({\cal D}_\alpha+{\cal A}_\alpha){\cal P}^{\beta\gamma}\ne 0$.

Heat kernel coefficients for non-minimal operators of this kind have been investigated recently 
\cite{Moss:2013cba}, although the results have to be extended slightly to include the explicit gauge
potential terms. In cases where ${\cal P}{\cal A}{\cal P}=0$, the traced heat kernel ${\rm Tr}\,E_1$
is a sum of invariants,
\begin{eqnarray}
{\rm Tr}\,E_1(\Delta_f)&=&a_1R\, {\rm Tr}(I)+a_2 {\rm Tr}\,{\cal M}^2+a_3 {\rm Tr}\,{\cal P}{\cal M}^2
+a_4R\,{\rm Tr}\,{\cal P}+a_5R\,{\rm Tr}\,{\cal P}^2\nonumber\\
&&+a_6t_{\mu\nu\rho\sigma}{\rm Tr}\,{\cal P}^{\mu[\nu}{\cal P}^{\alpha]\rho}{\cal P}^{\beta\sigma}F_{\alpha\beta}
+a_7{\rm Tr}\,{\cal A}_\mu{\cal A}^\mu+
a_8t_{\mu\nu\rho\sigma}{\rm Tr}\,{\cal P}^{\mu\nu}{\cal A}^\rho{\cal A}^\sigma,\label{tre1}
\end{eqnarray}
where ${\cal P}=g_{\mu\nu}{\cal P}^{\mu\nu}$,
$t_{\mu\nu\rho\sigma}=3g_{(\mu\nu}g_{\rho\sigma)}/m(m+2)$
and the coefficients are given in table \ref{table1}. 

\begin{table}[htb]
\caption{\label{table1}The coefficients of the invariants in the traced heat kernel coefficient 
${\rm Tr}\,E_1(\Delta)$ for operators with non-minimal term 
$-\zeta{\cal P}^{\mu\nu}{\cal D}_\mu{\cal D}_\mu$ in spacetime dimension
$m$, where $u=(1+\zeta)^{-m/2}$.}
\centering
\begingroup
\renewcommand{\arraystretch}{3.0}
\begin{tabular}{ll}
\hline\noalign{\smallskip}
Term&Expression\\
%\hline
\noalign{\smallskip}\hline\noalign{\smallskip}
$a_1$&$\displaystyle\frac16$\\
$a_2$&$-1$\\
$a_3$&$\displaystyle-\frac{u-1}{ m}$\\
$a_4$&$\displaystyle-{\frac { \left( m+2 \right)  \left( m\, \zeta  -
2\,m+4\,\zeta +10 \right) (u-1)}{ 12\left( m-2 \right) m\, \left( m-1
 \right) }}+{\frac {  \zeta  \, \left( -9\,m+{m}^{2
}+2 \right) }{6 \left( m-2 \right) m\, \left( m-1 \right) }}
$\\
$a_5$&$\displaystyle{\frac { \left( 8+{m}^{2} \zeta +4\,\zeta 
 \right) (u-1)}{4 \left( m-2 \right)  \left( m-1 \right) {m}^{2}}}+{\frac {
\zeta }{ \left( m-2 \right) m\, \left( m-1 \right) }}
$
\\
$a_6$&$\displaystyle{\frac { 2\left( 4+2\,\zeta +m\,  \zeta  \right) (u-1)}{m
-2}}+{\frac {4m\, \zeta }{m-2}}$
\\
$a_7$&$-1$
\\
$a_8$&$\displaystyle{\frac{4(u-1)}{\zeta}+2m}$
\\
\noalign{\smallskip}\hline
\end{tabular}
\endgroup
\end{table}

In the case of interest, the curvature 
$F_{\alpha\beta}$ is the curvature of the spin 2 tetrad connection, which is used in the 
derivatives of the metric fluctuations. Since the only terms which mix the spin 2 and scalar 
sectors are the terms involving ${\cal A}_\mu$, a simplification can be made by
spitting the result into a part depending on a spin 2 operator $\Delta^{(2)}$, a part 
depending on a scalar operator $\Delta^{(0)}$, and the cross terms,
rewriting ${\rm Tr}\,E_1$ as
\begin{equation}
{\rm Tr}\,E_1(\Delta_f)={\rm Tr}\,E_1(\Delta^{(2)})+{\rm Tr}\,E_1(\Delta^{(0)})
+a_7{\rm Tr}\,{\cal A}_\mu{\cal A}^\mu+
a_8t_{\mu\nu\rho\sigma}{\rm Tr}\,{\cal P}^{\mu\nu}{\cal A}^\rho{\cal A}^\sigma.
\end{equation}
The mass terms in the spin 2 and scalar sectors can be read off from the 
Lagrangian densities (\ref{Lg}-\ref{Lv}),
\begin{eqnarray}
{\cal M}^2{}_{\mu\nu}{}^{\rho\sigma}&=&
g_{(\mu\nu)(\alpha\beta)}\left\{
2\kappa^2g^{\alpha\rho}\partial^\beta\varphi^i\partial^\sigma\varphi_i
-2R^{\alpha\rho\beta\sigma}\right.\nonumber\\
&&\left.-\frac12g^{\alpha\beta}\left(\partial^\rho\varphi^i\partial^\sigma\varphi_i
-R^{\rho\sigma}\right)-\frac12g^{\rho\sigma}\left(\partial^\alpha\varphi^i\partial^\beta\varphi_i
-R^{\alpha\beta}\right)\right\}\nonumber\\
&&-{1\over m-2}\delta_\mu{}^{(\rho}\delta_\nu{}^{\sigma)}(G-\kappa^2 T),\label{m1}\\
{\cal M}^2{}_{i}{}^{j}&=&2(1+\zeta)\kappa^2\partial_\mu\varphi_i\partial^\mu\varphi^j
-{\cal R}_{ik}{}^j{}_l\partial_\mu\varphi^k\partial^\mu\varphi^l\nonumber\\
&&+V_{;i}{}^j-(m-2)^{-1}(G-\kappa^2 T)\delta_i{}^j\label{m2}
\end{eqnarray}
Some of the Vilkovisky-DeWitt corrections are evident in the $G-\kappa^2T$ terms, but
some have cancelled with other terms. The mass terms can be inserted into the results 
for ${\rm Tr}\,E_1$ with non-minimal spin 2 operators found in \cite{Moss:2013cba}.
Totalling all this together in $m=4$ dimensions and taking the Landau gauge $\zeta\to \infty$ limit gives
\begin{eqnarray}
\lim_{\zeta\to \infty}{\rm Tr}\,E_1(\Delta_f)&=&\
-{N+12\over 3}R+2(N+6)\kappa^2V\nonumber\\
&&+{N-4\over 2}\kappa^2\partial_\mu\varphi^i\partial^\mu\varphi_i
+{\cal R}_{ij}\partial_\mu\varphi^i\partial^\mu\varphi^j-V_{;i}{}^i.\label{vdwe1}
\end{eqnarray}
For comparison, the Feynman gauge result retaining the Vilkovisky-DeWitt corrections  is
\begin{eqnarray}
\left.{\rm Tr}\,E_1(\Delta_f)\right|_{\zeta=0}&=&\
-{N+13\over 3}R+2(N+10)\kappa^2V\nonumber\\
&&+{N-4\over 2}\kappa^2\partial_\mu\varphi^i\partial^\mu\varphi_i
+{\cal R}_{ij}\partial_\mu\varphi^i\partial^\mu\varphi^j-V_{;i}{}^i.\label{fge1}
\end{eqnarray}
The effect of changing the gauge fixing is quite small, but a large change comes 
about if we take the non-covariant  Feynman gauge result for the operator $\Delta^{\rm nc}$, 
defined by leaving out Vilkovisky-DeWitt corrections,
\begin{eqnarray}
\left.{\rm Tr}\,E_1(\Delta^{\rm nc})\right|_{\zeta=0}&=&\
{N-26\over 6}R+20\kappa^2V\nonumber\\
&&-2\kappa^2\partial_\mu\varphi^i\partial^\mu\varphi_i
+{\cal R}_{ij}\partial_\mu\varphi^i\partial^\mu\varphi^j-V_{;i}{}^i.\label{nce1}
\end{eqnarray}
The difference between ({\ref{fge1}) and (\ref{nce1}) is proportional to the field equation
$G-\kappa^2 T$ as expected.

The covariant result  (\ref{vdwe1}) is combined with the contribution from the ghost operator 
$\Delta_g$, which is of minimal type
\begin{equation}
\Delta_g=-\delta_\mu{}^\nu\nabla^2-R_\mu{}^\nu-2\kappa^2\partial_\mu\varphi^i\partial^\nu\varphi_i
\end{equation}
Only the first two terms in (\ref{tre1}) contribute to ${\rm Tr}\,E_1(\Delta_g)$,
\begin{equation}
{\rm Tr}\,E_1(\Delta_g)=\frac53R-2\kappa^2\partial_\mu\varphi^i\partial^\mu\varphi_i.\label{ge1}
\end{equation}
The quadratic divergence (\ref{qd}) can be obtained from the general expressions (\ref{dp}) and (\ref{gdiv}) 
using the heat kernel coefficients (\ref{vdwe1}) and (\ref{ge1}).

\section{The $E_2$ heat kernel coefficient}\label{appc}

The general expression for the traced heat kernel coefficient ${\rm Tr}\,E_2$ for the
non-minimal operator (\ref{op}) is unknown, but a simplified version with
${\cal A}_\mu=0$ can be obtained using the methods described in \cite{Moss:2013cba}.
The scalar and gravity sectors mix through the off-diagonal terms in the mass matrix.
If these are denoted by ${\cal M}^2_{\Delta}$, then the $E_2$ coefficient can be
expressed in terms of the results for the spin 2 and scalar parts,
\begin{equation}
{\rm Tr}\,E_2(\Delta_f)={\rm Tr}\,E_2(\Delta^{(2)})+{\rm Tr}\,E_2(\Delta^{(0)})
+c_1{\rm Tr}\,{\cal M}_{\Delta}^4+c_2{\rm Tr}\,{\cal P}{\cal M}_{\Delta}^4,
\end{equation}
where the coefficients are given in table \ref{table2}. The mass terms in the spin 2 and 
scalar sectors are given in (\ref{m1}) and (\ref{m2}), and the cross terms which 
define ${\cal M}^2_\Delta$ when $\partial_\mu\phi^i=0$ are
\begin{eqnarray}
{\cal M}^2{}_{\mu\nu}{}^j&=&-\frac12\kappa g_{\mu\nu}V^{;j},\label{ct1}\\
{\cal M}^2{}_i{}^{\rho\sigma}&=&\frac12\kappa g^{\rho\sigma}V_{;i}.\label{ct2}
\end{eqnarray}
The tensor $P^{\alpha\beta}$ is given in (\ref{ptens}).
The cross terms (\ref{ct1}) and (\ref{ct2}) can be regarded in Feynman diagram language as
scalar-scalar-graviton interaction vertices with one scalar from the background. The
Vilkovisky-DeWitt corrections have flipped the signs of these terms, but this does
not have any affect on the $E_2$ heat kernel coefficient.

\begin{table}[htb]
\caption{\label{table2}The coefficients of the invariants in the traced heat kernel coefficient 
${\rm Tr}\,E_2(\Delta)$ for operators in flat spacetime with non-minimal term 
$-\zeta{\cal P}^{\mu\nu}{\cal D}_\mu{\cal D}_\mu$ in spacetime dimension
$m$, where $u=(1+\zeta)^{-m/2}$.}
\centering
\begingroup
\renewcommand{\arraystretch}{3.0}
\begin{tabular}{llrr}
\hline\noalign{\smallskip}
Term&Expression&$\zeta=0$&$\quad\zeta\to\infty$\\
%\hline
\noalign{\smallskip}\hline\noalign{\smallskip}
$c_1$&$\displaystyle\frac12$&$\displaystyle\frac12$&$\displaystyle\frac12$\\
$c_2$&$\displaystyle-{2(1+\zeta)(u-1)\over \zeta m(m-2)}-{1\over m-2}$&$0$&$\displaystyle-\frac{1}{m}$\\
\noalign{\smallskip}\hline
\end{tabular}
\endgroup
\end{table}

The $E_2$ coefficients for spin 2 were evaluated in \cite{Moss:2013cba}. Taking the space-time dimensions $m=4$ 
and the Landau-gauge limit $\zeta\to\infty$
gives
\begin{eqnarray}
\lim_{\zeta\to\infty}{\rm Tr}\,E_2(\Delta_f)&=&\frac12V_{;ij}V^{;ij}
-\frac32\kappa^2V_{;i}V^{;i}-2\kappa^2VV_{;i}{}^i+(2N+12)\kappa^4 V^2\nonumber\\
&&-{2N+24\over3}\kappa^2RV+\frac13RV_{;i}{}^i+\frac{190+N}{180}R_{\mu\nu\rho\sigma}^2
+\frac{50-N}{180}R_{\mu\nu}^2+{41+2N\over 36}R^2.
\end{eqnarray}
For comparison, the non-covariant result obtained using Feynman gauge $\zeta=0$ 
would be
\begin{eqnarray}
\left.{\rm Tr}\,E_2(\Delta_f)\right|_{\zeta=0}&=&\frac12V_{;ij}V^{;ij}
-\kappa^2V_{;i}V^{;i}-2\kappa^2VV_{;i}{}^i+(2N+20)\kappa^4 V^2\nonumber\\
&&-{2N+26\over3}\kappa^2RV+\frac13RV_{;i}{}^i+\frac{190+N}{180}R_{\mu\nu\rho\sigma}^2
-\frac{190+N}{180}R_{\mu\nu}^2+{41+2N\over 36}R^2.
\end{eqnarray}
The Feynman-gauge curvature terms are consistent with previous work, e.g. 
Ref. \cite{barvinsky1985generalized}.
The ghost contribution is
\begin{equation}
{\rm Tr}\,E_2(\Delta_g)=-\frac{11}{180}R_{\mu\nu\rho\sigma}^2
+\frac{43}{90}R_{\mu\nu}^2+\frac29R^2.
\end{equation}
These combine to give the logarithmic divergences.

%%%%%%%%%%%%%%%%%%%%%%%%%%%%%%%%%%%%%%%%%%%%%
\bibliography{paper.bib}
\end{document}